\documentclass[fleqn,usenatbib]{mnras}

\usepackage{newtxtext,newtxmath}

\usepackage[T1]{fontenc}

\DeclareRobustCommand{\VAN}[3]{#2}
\let\VANthebibliography\thebibliography
\def\thebibliography{\DeclareRobustCommand{\VAN}[3]{##3}\VANthebibliography}

\usepackage{graphicx}
\usepackage{amsmath}
\usepackage{mathtools}
\usepackage{bm}
\usepackage{xcolor}
\usepackage{gensymb}
\newcommand{\rs}{\text{R}_\odot}
\newcommand{\dash}{\text{ -- }}
\newcommand{\bb}{\bm{B}}
\newcommand{\vb}{\bm{v}}
\newcommand{\VA}{V_\text{A}}
\newcommand{\MA}{M_\text{A}}
\newcommand{\Vsw}{V_\text{sw}}
\newcommand{\kmps}{\text{km s}^{-1}}

\newcommand{\psp}{\textit{PSP}}
\newcommand{\Rb}{\bm{\mathcal{R}}}

\title[Fragmented Alfv\'en Zone in Young Solar Wind]{An Extended and Fragmented Alfv\'en Zone in the Young Solar Wind}

\author[R. Chhiber et al.]{
Rohit Chhiber$^{1,2}$\thanks{E-mail: rohitc@udel.edu},
William H. Matthaeus$^1$,
Arcadi V. Usmanov$^{1,2}$,
Riddhi Bandyopadhyay$^3$, \newauthor
and Melvyn L. Goldstein$^4$
\\ \\
$^{1}$Department of Physics and Astronomy, University of Delaware, Newark, DE 19716, USA\\
$^{2}$Heliophysics Science Division, NASA Goddard Space Flight Center, Greenbelt, MD 20771, USA\\
$^{3}$Department of Astrophysical Sciences, Princeton University, Princeton, NJ 08544, USA\\
$^4$University of Maryland Baltimore County, Baltimore, MD 21250, USA
}

\date{Accepted XXX. Received YYY; in original form ZZZ}

\pubyear{2022}

\begin{document}
\label{firstpage}
\pagerange{\pageref{firstpage}--\pageref{lastpage}}
\maketitle

\begin{abstract}
Motivated by theoretical, numerical, and observational 
evidence, we explore the possibility that the critical transition between sub-Alfv\'enic flow and super-Alfv\'enic flow in the solar atmosphere takes place in fragmented and disconnected subvolumes within a general Alfv\'en critical zone. 
The initial observations of sub-Alfv\'enic periods by Parker Solar Probe near \(16~\rs\)  do not yet provide sufficient  evidence to distinguish this possibility from that of
a folded surface that separates simply-connected regions. Subsequent orbits may well enable such a distinction, but here we use a global magnetohydrodynamic model of the solar wind, coupled to a turbulence transport model, to generate possible realizations of such an Alfv\'en critical zone. Understanding this 
transition will inform theories of coronal heating, solar wind origin, solar angular momentum loss, and related physical processes in stellar winds beyond the Sun.  
\end{abstract}

\begin{keywords}
Sun: corona -- solar wind -- turbulence
\end{keywords}



\section{Introduction}
Based 
on  known variations of 
magnetic field, density, and solar wind flow speed, there  has long been an expectation of a transition between lower coronal sub-Alfv\'enic wind 
and super-Alfv\'enic solar wind \citep{weber1967ApJ148}. 
In particular, given a solar wind speed 
\(\Vsw\)\ and an Alfv\'en speed \(\VA\), 
one expects that a sub-Alfv\'enic state 
\(\Vsw < \VA\) exists close to the 
sun, 
while a super-Alfv\'enic
state \(\Vsw > \VA\) would dominate near 
Earth orbit.
In the simplest wind models the transition occurs at 
an  Alfv\'enic critical point. 
In three dimensions this
readily generalizes to a smooth critical surface \citep[e.g.,][and references within]{chhiber2019psp1}. 
Identification of this Alfv\'enic transition has been of interest for NASA's Parker Solar Probe (PSP) mission since the planning stages \citep{matthaeus2004_alfven_region,fox2016SSR}. In fact, it has already been suggested \citep{matthaeus2004_alfven_region,deforest2018ApJ,chhiber2019psp1,wexler2021ApJ} that 
the transition might 
be better described as a trans-Alfv\'enic \textit{region} or 
\textit{zone}, rather than a simple surface,
in part based on remote sensing observations of an ``extended solar wind
transsonic (sic) region'' \citep{lotova1985AA150,lotova1988SoPh117}. 

Even if the corona formally refers to the
entirety of the solar atmosphere, there 
is some utility in drawing the 
distinction between the 
magnetically dominated sub-Alfv\'enic region as the
pure corona, and the extended super-Alfv\'enic flow-dominated 
wind that permeates interplanetary space. Indeed, the crucial processes that heat the corona and accelerate the wind originate in the 
magnetically dominated region, thus providing a fundamental physics motivation for the PSP mission \citep{fox2016SSR}.
Recently, on its eighth perihelion, at a distance of about \(19~\rs\), 
PSP passed into sub-Alfv\'enic plasma for extended periods for the first time 
\citep{kasper2021prl}.
This has enabled the first glimpse of the 
magnetically dominated coronal plasma using 
{\it in situ} observations. The leading 
order expectations of distinctive features of coronal 
plasma and electromagnetic fields thus far appear to be intact \citep{kasper2021prl,BandyopadhyayEA21}.
In the present paper we explore 
how subsequent passages through this sub-Alfv\'enic coronal plasma 
might look in subsequent PSP orbits
if, indeed, 
the trans-Alfv\'enic 
boundary is a zone rather
than a well-defined surface. 

The first passages of PSP into the sub-Alfv\'enic region have 
comprised several time periods, the longest of which is a few hours \citep{kasper2021prl}. 
Further observations in more persistent sub-Alfv\'enic 
wind will be required for confirmation of these early measurements. What remains unclear is the 
topological nature of the trans-Alfv\'enic 
zone. Single-point observations such as those made by PSP
do not readily distinguish between
a folded but smooth Alfv\'en surface, a highly corrugated surface, 
or an even more complicated patchy, disconnected, or even fractal surface. Here we will 
develop a
hybrid model -- 
part global magnetohydrodynamic (MHD) 
and part 
synthetic and empirically justified, 
to visualize 
what an extended, 
fragmented Alfv\'en zone may appear like. Understanding this possibility in some detail may help to identify the morphology
of the Alfv\'enic transition based on subsequent PSP orbits.  

The Alfv\'en surface departs from a simple radial critical point in even moderately realistic three-dimensional (3D) models such as source surface mappings and MHD simulations 
\citep[e.g.,][]{cohen2015SoPh,chhiber2019psp1}.
When based on magnetogram boundary conditions, 
these models suggest a critical surface that exhibits asymmetries in both latitude and longitude, with significant inward
distortion near the heliospheric current sheet (HCS).
Furthermore, 
recent papers based on extrapolation of in-situ measurements infer a ``rugged'' Alfv\'en surface 
\citep{wexler2021ApJ,liu2021ApJ,Verscharen2021MNRAS}

It is also expected that when 
fluctuations of sufficient amplitude are included, 
the Alfv\'en surface will become even 
more complex. Recall that the Alfv\'en Mach number is defined 
as 
\begin{equation}
    M_\text{A}({\bf r}) = \frac{\Vsw}{\VA} = 
    \frac{\Vsw({\bf r})} {B({\bf r})/\sqrt{4 \pi \rho({\bf r})}}
    \label{eq:MA}
\end{equation}
where \(\Vsw\), the magnetic field strength $B$, and the mass density $\rho$ are all functions of position.
Now suppose that the total speed, magnetic 
field, and density contain 
fluctuating components, which may be indicated as 
$\Vsw = \langle\Vsw\rangle + \Vsw^\prime$, 
$\bm{B} = \langle\bm{B}\rangle +  \bm{B}^\prime$ and $\rho = \langle\rho\rangle + \rho^\prime$,
where the \(\langle\dots\rangle\) indicates the mean value and the prime indicates a fluctuating component that averages to zero over a few
correlation scales.
It is clear that when such fluctuations are present, there will be corresponding fluctuations of the Alfv\'en Mach number. 

There are, in fact,
a variety of reasons to 
expect that fluctuation amplitudes increase moving with the solar wind toward the Alfv\'en transition. 
Not only are there 
signatures of their presence in 
lower solar atmospheric observations \citep{Tomczyk2007Sci,Mathioudakis2013SSR}, 
but this is also expected even on the basis of WKB theory \citep{hollweg1973JGR,velli1993AA}.
Turbulence transport theory, containing a generalization of 
WKB theory, predicts an 
increase in fluctuation amplitude in the expanding 
coronal plasma, peaking near the Alfv\'en zone \citep{verdini2010ApJ}. 
When combined self-consistently
with a large-scale global solar wind MHD model
\citep{usmanov2011solar,usmanov2012three,usmanov2014three,usmanov2018}, the transport theory
predicts a strong maximum of fluctuation amplitude near the Alfv\'en critical zone \citep{chhiber2019psp1}. Such results can also be reproduced by an expanding box model \citep{squire2020ApJ}
that contains physics similar to 
that of the turbulence transport models \citep[see also][]{zank2021PoP}.

Interestingly, there have also been suggestions of a strong turbulence region near 10-30 \(\rs\) based on remote-sensing observations \citep{lotova1985AA150,lotova1988SoPh117,lotova1997SoPh172}. Additional evidence of large fluctuations, in this case in velocity, is provided 
by highly sensitive chronograph observations \citep{deforest2018ApJ}. In the latter case remote sensing indicates the presence of  radially flowing electron density blobs moving outward along stream tubes with differential speeds of approximately \(200~\kmps\)\ along neighbouring stream tubes.  

Given these diverse expectations as to the amplitudes reached 
by fluctuations near 
the Alfv\'en critical zone, it is 
understandable, 
in accord with Equation \ref{eq:MA},
that the Alfv\'en Mach number itself will fluctuate along a stream tube. 
Furthermore, if fluctuation amplitudes peak near the Alfv\'en zone, fluctuations 
of the Alfv\'en Mach number itself
will peak in that region. Based on this reasoning, the critical Alfv\'en ``surface'' should be neither smooth nor a simply folded surface that each streamline passes though only once.
Here we explore the possibility that there are regions or subvolumes of plasma that exist in a sub-Alfv\'enic state intermixed with subvolumes of super-Alfv\'enic flow. 
Thus, 
along flow tubes of solar wind 
one would repeatedly encounter subregions
of sub/super-Alfv\'enic 
solar wind flows. 
Accordingly,
the present paper is devoted to 
realizations of the Alfv\'en critical zone that 
include the effects of self consistently computed fluctuations. In this way we can 
quantitatively assess features of the critical zone 
that may potentially 
contribute to our understanding of
physical processes in the solar wind. 
We expect that the validity of this characterization
of a fragmented Alfv\'en zone 
will be either corroborated or 
invalidated
in future PSP orbits, as well as in
anticipated observations
from next-generation remote imaging missions such as PUNCH \citep{deforest2019AGU_PUNCH}. 

In outline, we briefly describe the solar wind model employed in this study in Section \ref{sec:theory}. Results are presented in Section \ref{sec:results}, which includes a comparison of model output with PSP observations. We conclude with a discussion in Section \ref{sec:disc}.

\section{Solar Wind Model with Turbulence Transport}\label{sec:theory}

Realistic and accurate MHD
coronal simulations that resolve all scales 
from the large features at the solar surface to the 
dissipation range of solar wind turbulence 
are excessively demanding computationally and are at present intractable \citep[e.g.,][]{miesch2015SSR194}. 
Direct simulations of turbulence 
effects on the Alfv\'en critical zone are included
in this class. 
Therefore, to proceed we adopt an approach based on Reynolds averaging \citep[e.g.,][]{mccomb1990physics} that 
provides explicit treatment of large scales and a self-consistent 
statistical treatment of the turbulence. 
We then generate realizations of turbulence
that, when combined with the large-scale solutions, produce a picture of a turbulence-modified Alfv\'en zone. 

A two-fluid magnetohydrodynamics (MHD)
model \citep{usmanov2018,gombosi2018LRSP} 
reasonably
describes the large-scale
features of the solar wind
when the internal energy 
is separated into electron and proton fluid ingredients \citep{cranmer2009ApJ}. The dynamics of these large-scale features is determined to a significant degree, but not completely, by 
boundary conditions, so that 
information flow is mainly along characteristics.  
Features separated in angle by more than a few tens of degrees do not communicate well \citep[see, e.g.,][]{matthaeus1986prl}. 
At smaller scales the system is subject to local turbulence interactions, which, although formally deterministic, are conveniently approximated by a statistical treatment, such as Reynolds averaging.

The two-fluid MHD global coronal and  
model that we employ is briefly summarized here, and is described in detail in \cite{usmanov2014three,usmanov2018}. The Reynolds-averaging approach is based on the decomposition
of physical fields, e.g., $\tilde{\bm{a}}$
into a mean and a fluctuating component: \(\tilde{\bm{a}} = \bm{a}+\bm{a}'\), making use of an ensemble-averaging operation where \(\bm{a} = \langle \tilde{\bm{a}} \rangle\) and, by construction, \(\langle\bm{a}'\rangle=0\). Application of this 
decomposition to the 
primitive compressible MHD equations,
along with a series of approximations appropriate to the solar
wind, leads to a set of mean-flow equations that are coupled
to the small-scale fluctuations via another set of equations for statistical descriptors of the unresolved
turbulence.

To derive the mean-flow equations, the velocity and magnetic fields are Reynolds decomposed: $\tilde{\bm{V}} = \bm{V}+\bm{v'}$ and $\widetilde{\bm{B}} = \bm{B}+\bm{B'}$, and then substituted into the momentum and induction equations in the frame of reference corotating with the Sun. The ensemble averaging operator $\langle\dots\rangle$ is then applied to these two equations \citep{usmanov2014three,usmanov2018}. The resulting mean-flow model consists of a single momentum equation and separate ion and electron temperature equations, in addition to an induction equation. 
Density fluctuations are neglected, and pressure
fluctuations are only those required to maintain incompressibility \citep{zank1992waves}. The Reynolds-averaging procedure introduces additional terms in the mean flow equations, 
representing the influence of turbulence on the mean (average) dynamics. These terms involve the Reynolds stress tensor \(\Rb = \langle\rho\vb'\vb' -
\bb'\bb'/4\pi\rangle\), the mean turbulent electric field
\(\mathcal{\bm{\varepsilon}}_m = \langle\vb'\times\bb'\rangle(4\pi\rho)^{-1/2}\), the
fluctuating magnetic pressure \(\langle B'^2\rangle/8\pi\), and the turbulent heating, or ``heat function'' \(Q_T(\bm{r})\), which 
is apportioned between protons and electrons. Here the mass density \(\rho=m_p N_S\) is defined in terms of the proton mass \(m_p\) and number density \(N_S\). The pressure equations employ the natural ideal gas 
value of 5/3 for the adiabatic index.
The pressure equations 
also include weak proton-electron collisional friction terms involving a classical Spitzer collision time scale
\citep{spitzer1965,hartle1968ApJ151}
to model the energy exchange between the
protons and electrons
\citep[see][]{breech2009JGRA}. Electron heat flux below \(5\text{ -- }10~\rs\) is approximated by the classical collision-dominated model of \cite{spitzer1953PhRv} \citep[see also][]{chhiber2016solar}, while above \(5 \text{ -- } 10~\rs\) we adopt Hollweg's ``collisionless'' 
conduction model \citep{hollweg1974JGR79,hollweg1976JGR}. We neglect proton
heat flux. See \cite{usmanov2018} for more details. 

Transport equations for the fluctuations, assumed to be at relatively small scales, 
are obtained by subtracting the mean-field equations from the full MHD equations and averaging the difference  \citep[see][]{usmanov2014three}. This yields 
equations \citep{breech2008turbulence,usmanov2014three,usmanov2018}
for the three chosen statistical descriptors of turbulence, namely \(Z^2 = \langle v'^2 + b'^2\rangle\), i.e., 
twice the fluctuation energy
per unit mass, where \(\bm{b}' = \bb'(4\pi\rho)^{-1/2}\);
the normalized cross helicity, 
or normalized cross-correlation between velocity and magnetic field fluctuations \(\sigma_c
= 2\langle\vb'\cdot\bm{b}'\rangle/Z^2\);
and $\lambda$, a correlation
length perpendicular to the mean magnetic field. Note that the assumption of a single correlation scale \citep[cf.][]{zank2018ApJ} implies structural similarity of autocorrelation functions of the turbulent fields; this assumption was found to be reasonably valid for PSP observations \citep{chhiber2021ApJ_psp}. 
Other parameters include the normalized energy difference, which we treat as a constant parameter (\(=-1/3\)) derived from
observations \citep[cf.][]{zank2018ApJ}, and the K\'arm\'an-Taylor
constants \cite[see][]{matthaeus1996jpp,smith2001JGR,
breech2008turbulence}. Note that the fluctuation energy loss due to 
our assumption of von K\'arm\'an decay \citep{karman1938prsl,hossain1995PhFl,
wan2012JFM697,wu2013prl,bandyopadhyay2018prx}  is balanced in a quasi-steady state by internal energy supply in the 
pressure equations. 
The Reynolds stress is simplified 
by assuming 
that the turbulence is transverse to the mean field and axisymmetric about it \citep{oughton2015philtran}, so that we obtain \(\Rb/\rho = K_R(\bm{I} - \hat{\bm{B}}\hat{\bm{B}})\), where \(K_R = \langle v'^2 - b'^2\rangle/2 = \sigma_D Z^2/2\) is the residual energy, and \(\hat{\bm{B}}\) is a unit vector in the direction of \(\bm{B}\). The turbulent electric field is neglected here. For further details see \cite{usmanov2018}.


We solve the Reynolds-averaged mean-flow equations concurrently with the turbulence transport equations in the spherical shell between the base of the solar corona (just above the transition region) and the heliocentric distance of 5 AU. The computational domain is split into two regions: the inner (coronal) region of 
\(1\dash 30~\rs\) and the outer (solar wind) region from \(30~\rs\dash 5\) AU. The relaxation method, i.e., the integration of time-dependent equations in time until a steady state is achieved, is used in both regions. The simulations have a resolution of \(702\times 120\times 240\) grid points along \(r\times \theta\times \phi\) coordinates. The computational grid has logarithmic spacing along the heliocentric radial (\(r\)) direction, with the grid spacing becoming larger as \(r\) increases. The latitudinal (\(\theta\)) and longitudinal (\(\phi\)) grids have equidistant spacing, with a resolution of 1.5\degree~each. In terms of physical scales, the grid spacing corresponds to several correlation lengths of magnetic fluctuations \citep[e.g.,][]{Ruiz2014SoPh}, thus providing strong motivation for the 
statistical model we employ for unresolved, sub-gridscale turbulence in the present study. 

The model is well-tested and has been shown to yield good agreement with a variety of observations \citep{breech2008turbulence,usmanov2011solar,usmanov2012three,usmanov2014three,chhiber2017ApJS230,usmanov2018,chhiber2018apjl,chhiber2019psp1,bandyopadhyay2020ApJS_cascade,ruffolo2020ApJ,chhiber2021ApJ_psp}.

\section{Results}\label{sec:results}

We now summarize the procedure 
employed to obtain the results below:
\begin{enumerate}
    \item Initialize the global heliospheric code \citep{usmanov2018} with magnetogram-based boundary conditions appropriate for a specified PSP solar encounter, or a selected tilted magnetic dipole boundary condition;
    \item Obtain Reynolds-averaged large-scale fields and statistical turbulence parameters from the model computation; 
    \item Employ self consistently determined turbulence parameters from the model to synthesize a plausible realization of the local turbulence fields;
    \item Combine mean fields and synthesized fluctuations to examine transitions between super-Alfv\'enic and sub-Alfv\'enic conditions in the trans-Alfv\'enic zone.
\end{enumerate}


\subsection{Model Runs \label{sec:runs}}

For the present study we performed three runs: Run I is based on a source magnetic dipole tilted by 10\degree~ relative to the solar rotation axis, toward 330\degree\ longitude in Heliographic Coordinates \citep[HGC;][]{franz2002pss}; Run II is based on a Wilcox Solar Observatory (WSO) magnetogram for Carrington Rotation 2215, corresponding to PSP's first solar encounter; Run III is based on an ADAPT map with central meridian time 2021 April 29 at 12:00 UTC, corresponding to PSP's eighth solar encounter (when the spacecraft first sampled sub-Alfv\'enic wind). The synoptic magnetograms from WSO have 5\degree\ resolution in heliolongitude
and 30 points equidistantly distributed over the sine of heliolatitude. The ADAPT maps, which are based on the GONG magnetogram \citep[][]{arge2010AIPC}, have a 1\degree-resolution both in heliolatitude and heliolongitude. The WSO and ADAPT magnetograms are scaled by a factor of 8 and 2, respectively\footnote{This scaling is required to obtain agreement between model results and spacecraft observations near Earth \citep[see][]{riley2014SoPh}. The choice of scaling factor and its effects on model output are discussed in detail by \cite{usmanov2018}.}, and smoothed using a spherical harmonic expansion up to \(9^\text{th}\) and \(15^\text{th}\) order, respectively. Input parameters specified uniformly at the coronal base include the
driving amplitude of Alfv\'en waves (30 km~s$^{-1}$), and the correlation
scale of turbulence (\(10,500\)~km). In the initial state, the density and
temperature are also prescribed uniformly ($8 \times 10^7$ particles
cm$^{-3}$ and $1.8 \times 10^6$~K, respectively), but they can change in
the course of relaxation to a steady state. The cross helicity in the initial state is set as \(\sigma_c = -\sigma_{c0} B_r/B_r^\text{max}\), where  \(\sigma_{c0}=0.8\), \(B_r\) is the radial magnetic field, and \(B_r^\text{max}\) is the maximum absolute value of \(B_r\) on the inner boundary. The input parameters also include the fraction of turbulent energy absorbed by protons $f_p = 0.6$, the normalized energy difference \(\sigma_D=-1/3\), and  K\'arm\'an-Taylor constants \(\alpha = 2\beta = 0.128\). Further details on the numerical approach and initial and boundary conditions may be found in \cite{usmanov2018}, who also examined the effects of varying these parameters on the model results.

\subsection{Statistical Boundaries of the Alfv\'en Zone}\label{sec:stat_zone}

\begin{figure*}
    \centering
    \includegraphics[width=1\columnwidth]{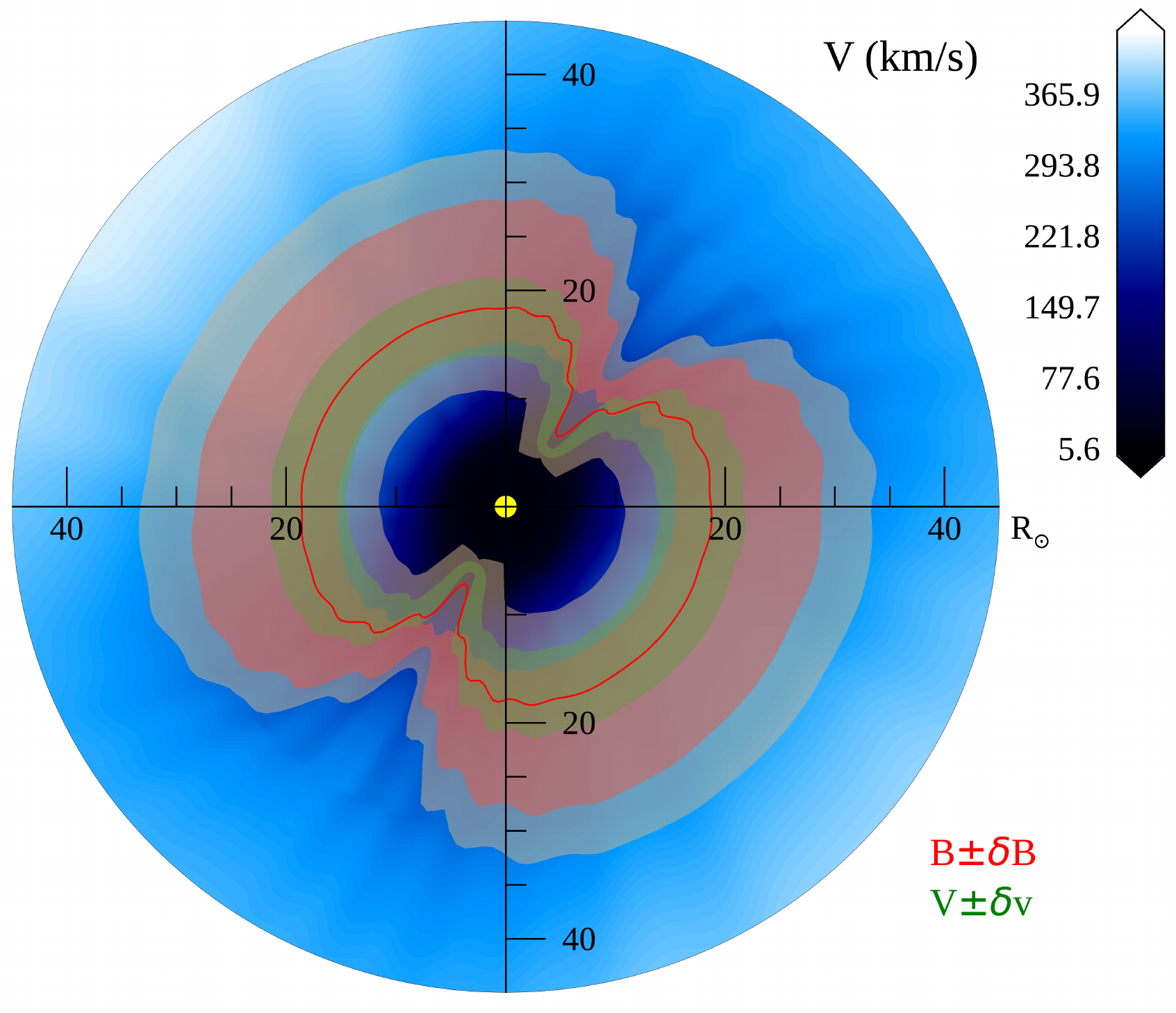}
    \includegraphics[width=1\columnwidth]{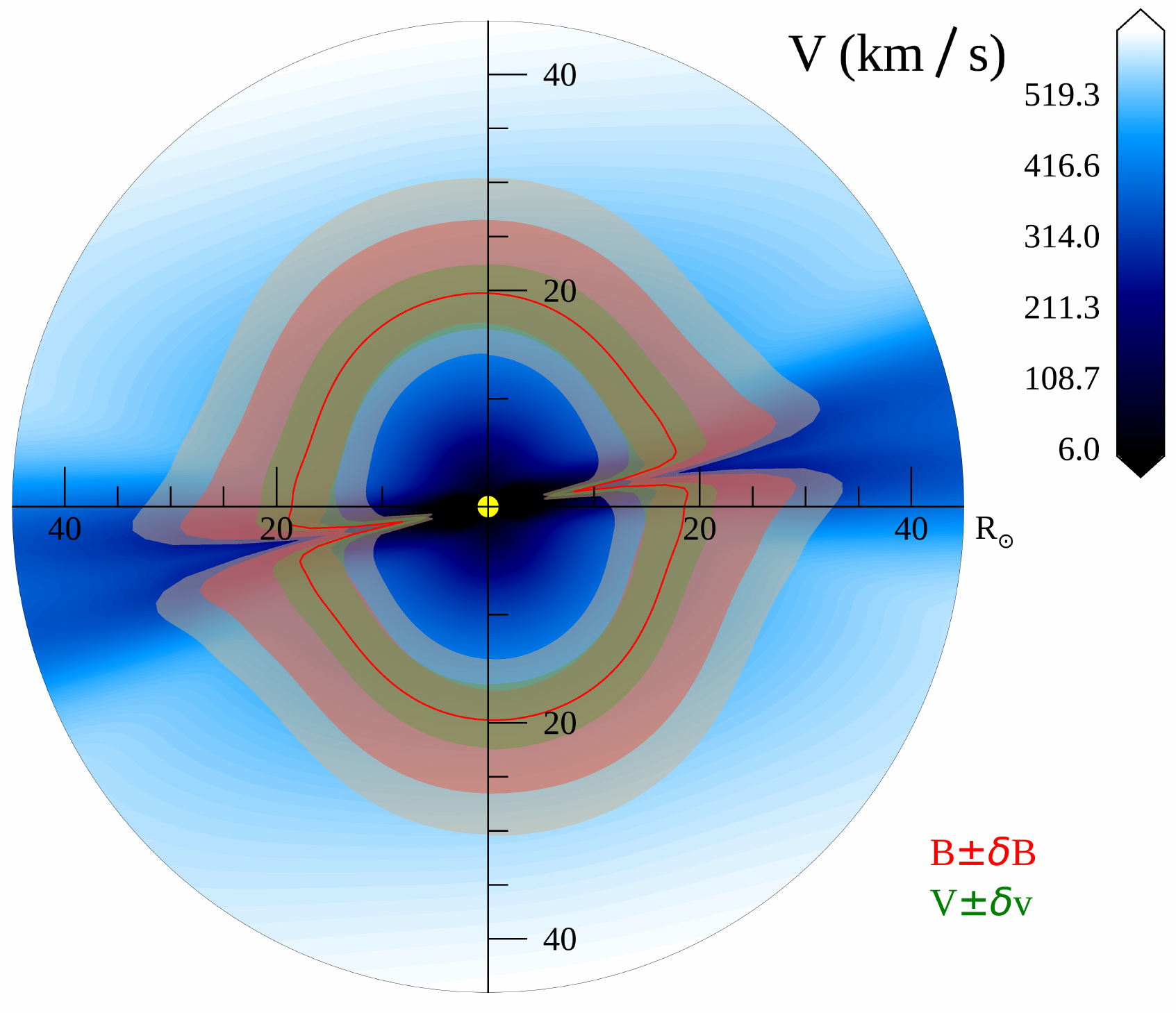}
    \caption{Solar wind speed and the effect of fluctuations on the Alfv\'en zone. Left and right panels show the solar equatorial plane (\(r\)-\(\phi\) plane at 0\degree~latitude) and a meridional plane (\(r\)-\(\theta\) plane at 155\degree~longitude), respectively, for the 10\degree~dipole-based Run I. Helioradii ranging from 1 to \(45~\rs\)~are shown in both panels. Red curve shows the Alfv\'en surface computed using the mean velocity and magnetic fields. Red-shaded region shows the range of distances at which the trans-Alfv\'enic transition can occur when magnetic fluctuations are accounted for (see text). Green-shaded region shows this trans-Alfv\'enic zone when  velocity fluctuations, but not magnetic fluctuations are included. The widest, beige-shaded region shows this zone when both magnetic and velocity fluctuations are included.}
    \label{fig:stat_zone}
\end{figure*}

Once the boundary condition is selected and the model parameters set, steps (i) and (ii) above are straightforward. Results of this type have been used to trace critical surfaces for comparison with remote imaging \citep{chhiber2018apjl,chhiber2019psp1}, to make predictions for PSP crossings of critical surfaces \citep{chhiber2019psp1}, and for comparison of model results with mean-field and turbulence measurements along the PSP trajectory for orbits 1 through 5 \citep{chhiber2021ApJ_psp}. 
Determination of a consistent 
realization of the turbulence in step (iii) 
is a novel procedure that we describe here
in order to estimate turbulence effects 
on the topography of the trans-Alfv\'enic critical region. 

To extract rms amplitudes for magnetic and velocity fluctuations from the \(Z^2\) variable computed in the turbulence transport model, 
we assume an Alfv\'en ratio \(r_\text{A} \equiv \langle v'^2\rangle/\langle B'^2\rangle = 0.5\), following observations from PSP and near-Earth spacecraft \citep{tu1995SSRv,chen2020ApJS,parashar2020ApJS}; this is also consistent with the constant energy difference \(\sigma_\text{D}=(r_\text{A}-1)/(r_\text{A}+1) = -1/3\) assumed in our model \citep[cf.][]{zank2018ApJ}. Then we obtain 3D distributions of rms velocity and magnetic fluctuation amplitudes using the expressions \(\delta v \equiv \langle v'^2\rangle^{1/2} = [Z^2/(1+1/r_\text{A})]^{1/2}\) and \(\delta B\equiv \langle B'^2\rangle^{1/2} = [Z^2 4 \pi\rho/(1+r_\text{A})]^{1/2}\), respectively. 

An analogous procedure is used to obtain the partitioning of fluctuation energy among the three local Cartesian components of polarization. The transport model described in Section \ref{sec:theory} adopts the approximation that the fluctuations have purely transverse polarizations, for analytical tractability \citep{breech2008turbulence,usmanov2014three}. However, to preserve the magnetic field's divergenceless nature, Alfv\'enic-type fluctuations of finite amplitude must include three components of polarization \citep{barnes1976JGR,barnes1979inbook}. We adopt a simple isotropic partitioning in this initial study, noting that the Alfv\'en Mach number is computed using the magnitudes of the solar wind speed and magnetic field, and is therefore not very sensitive to the relative strengths of individual local Cartesian components. Then each fluctuating component \(\delta B_i\) has a variance \(\delta B_i^2 = \delta B^2/3\), where \(i \in \{r,\theta,\phi\}\). A similar procedure is used to estimate components of velocity fluctuations.

With these preparations we are now in a position to examine the effect of turbulence on the location of the Alfv\'en surface.
We estimate upper and lower statistical bounds for the total magnetic and velocity fields, taking into account rms fluctuations about the mean values: the upper and lower bounds for the solar wind speed are computed as \(V_\pm = |\bm{V}\pm \delta \bm{v}| = [(V_r\pm \delta v_r)^2 + (V_\theta\pm \delta v_\theta)^2 + (v_\phi\pm \delta v_\phi)^2]^{1/2}\), and the corresponding bounds for the Alfv\'en speed are \(V_{\text{A}\pm} =  [(B_r\pm \delta B_r)^2 + (B_\theta\pm \delta B_\theta)^2 + (B_\phi\pm \delta B_\phi)^2]^{1/2} /(4\pi\rho)^{1/2}\). 

Based on these procedures,
Figure \ref{fig:stat_zone} 
illustrates features of 
the Alfv\'en critical zone 
represented in the solar equatorial plane (left panel) and a meridional plane (right panel). The background shows a colour map of the solar wind speed, exhibiting familiar features of acceleration and fast/slow wind streams \citep{mccomas2003grl}. The trans-Alfv\'enic boundary is computed in four different ways: The red curve shows the Alfv\'en surface computed in the conventional way \citep[e.g.,][]{chhiber2019psp1}, as the set of points where the mean solar wind speed \(V\) first becomes larger than the Alfv\'en speed computed from the mean magnetic field: \(V_A = B/(4\pi\rho)^{ 1/2}\). Away from the HCS this ``mean'' Alfv\'en surface is located at \(\sim 18~\rs\). The red-shaded Alfv\'en zone is bound by outer and inner envelopes defined by the set of points where the condition \(V > V_{\text{A}\pm}\) is first attained approaching from
larger radial distances (outer envelope),
or from smaller radial distances (inner envelope); only the effect of magnetic fluctuations is taken into account. Note that the mean-field Alfv\'en surface marked by the red curve dips below the red-shaded zone at the HCS, where the mean magnetic field vanishes, leading to a small mean Alfv\'en speed. The green-shaded zone is obtained in a similar way, by considering the condition \(V_\pm > V_\text{A}\), i.e., accounting for velocity fluctuations only. Finally, the beige-shaded zone accounts for both magnetic and velocity fluctuations by using the condition \(V_\pm > V_{\text{A},\pm}\).
One may observe from this relatively crude exercise that, away from the HCS, and for this particular case, 
the position of the critical surface can be 
displaced, on average, by 5 to 10 \(~\rs\) by the (approximate) influence of magnetic turbulence. Velocity fluctuations produce a relatively weaker effect (partly due to the choice of Alfv\'en ratio), which is also symmetric about the ``mean'' Alfv\'en surface (red curve). As expected, combining velocity and magnetic fluctuations produces the largest variability, with the Alfv\'en surface shifted by as much as \(15~\rs\).

\subsection{Realization of a Fragmented Alfv\'en Zone}
\label{sec:frag}

While the above procedure permits estimation of the \textit{average} spatial extent of the Alfv\'en critical zone, 
we can provide additional detail, 
again in an approximate but consistent sense, 
by adopting a specific field of turbulence that is based on the turbulence model 
incorporated in the Reynolds-averaged MHD simulations. 
To examine the effect of an explicit realization of turbulence on the Alfv\'en zone, we generate synthetic random fluctuations that are constrained by the rms turbulence amplitudes described in the previous section.\footnote{A similar approach was used in \cite{chhiber2021ApJ_psp} for a comparison of synthetic magnetic fluctuations with PSP observations.} 
These fluctuations, inserted into 
Equation \ref{eq:MA}, enable us to examine 
a plausible 
spatial dependence of the
Alfv\'en Mach number while including the effect of turbulence.

For this first study, 
we examine the effect of magnetic fluctuations, 
although velocity and density fluctuations
can also, in principle,
be incorporated in an 
analogous way. 
At each point on the simulation grid, \(\{r,\theta,\phi\}\) components of a random magnetic fluctuation vector are generated: each component \((B'_r, B'_\theta, B'_\phi)\) is a random number obtained from a Gaussian distribution with standard deviation equal to the respective rms magnetic fluctuation component given by the model at that point (see Section \ref{sec:stat_zone}).\footnote{We use the IDL function {\fontfamily{qcr}\selectfont \href{https://www.l3harrisgeospatial.com/docs/randomu.html}{randumu}}, which is based on the Mersenne Twister algorithm for generating pseudo-random numbers \citep{matsumoto1998ACM}. Note that these numbers may have positive or negative sign.} The total magnetic field magnitude is then \(\tilde{B} = [(B_r + B_r^\prime)^2 + (B_\theta + B'_\theta)^2 + (B_\phi + B'_\phi)^2]^{1/2}\), and thus we obtain a 3D distribution of Alfv\'en speed that includes an explicit realization of magnetic fluctuations: \(\tilde{V}_\text{A}=\tilde{B}/\sqrt{4\pi \rho}\). In the following, \(\tilde{V}_\text{A}\) is compared with the mean solar wind speed \(V\) to compute the local Alfv\'en Mach number at a simulation grid point.

\begin{figure}
    \centering
    \includegraphics[width=1\columnwidth]{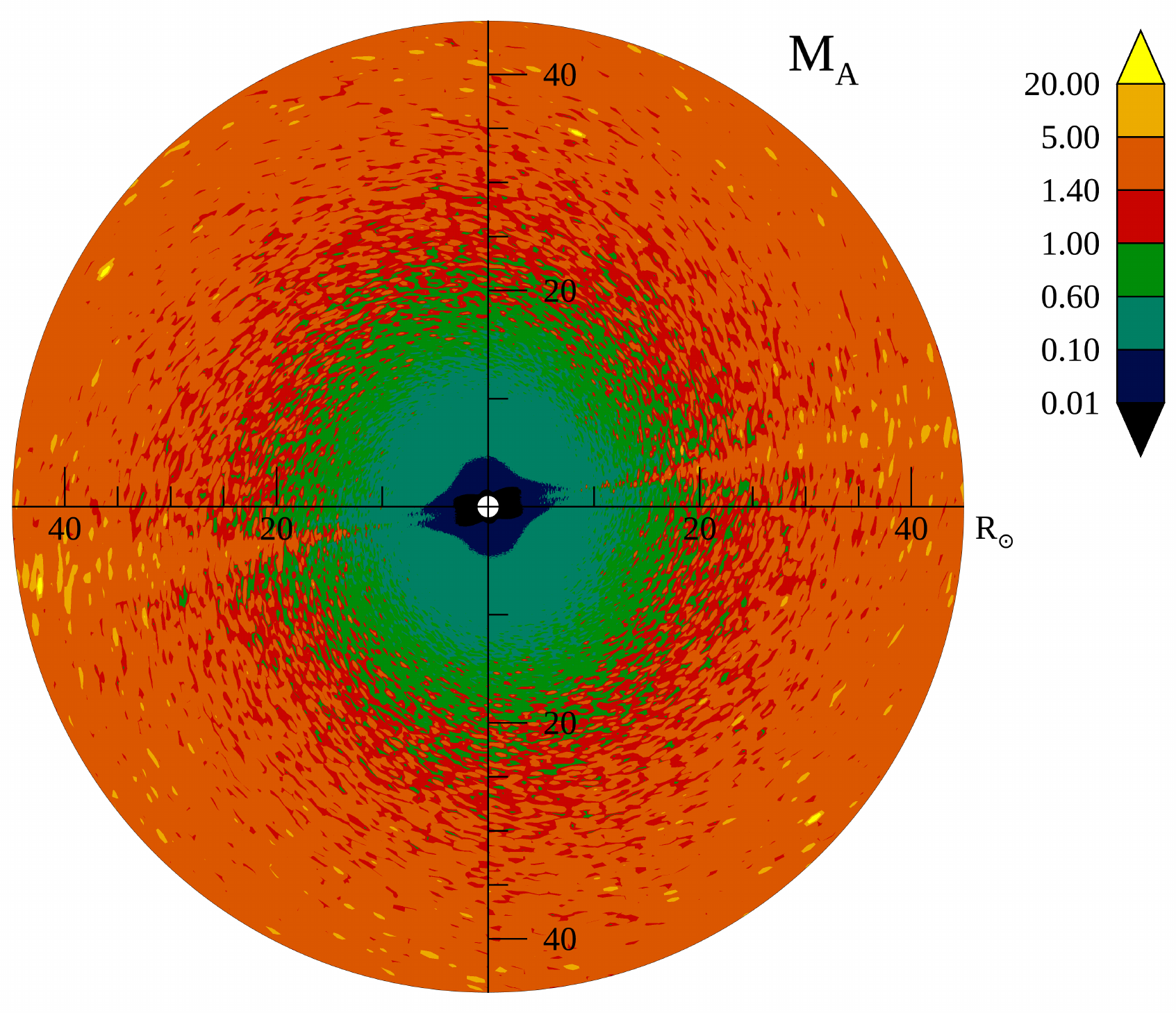}
    \caption{Alfv\'en Mach number in a meridional plane at 155\degree~longitude, from Run I (\(10\degree\)-tilt dipole). Helioradii ranging from 1 to \(45~\rs\)~are shown. The Alfv\'en speed is computed from a magnetic field that includes an explicit realization of fluctuations, constrained by the turbulence transport model (see text).}
    \label{fig:MA_merid_10tilt}
\end{figure}
Figure \ref{fig:MA_merid_10tilt} depicts the values of $M_\text{A} = V/\tilde{V}_\text{A}$ in a meridional plane
for a \(10\degree\)-tilt dipole simulation (Run I). The image is nearly rotationally symmetric except for the indentation feature associated with the HCS. With the exception of that region, the many transitions from low Alfv\'en Mach number (shades of green/black) to high Alfv\'en Mach number 
(shades of red/orange) occur mainly in a band that extends from around 
12 $\rs$ to about 25 $\rs$. The range of this band is comparable to what is seen in the red-shaded region of Figure \ref{fig:stat_zone}. Within this band we observe patches of plasma with \(\MA >1\) interspersed with sub-Alfv\'enic patches.

\begin{figure*}
    \centering
    \includegraphics[width=1\columnwidth]{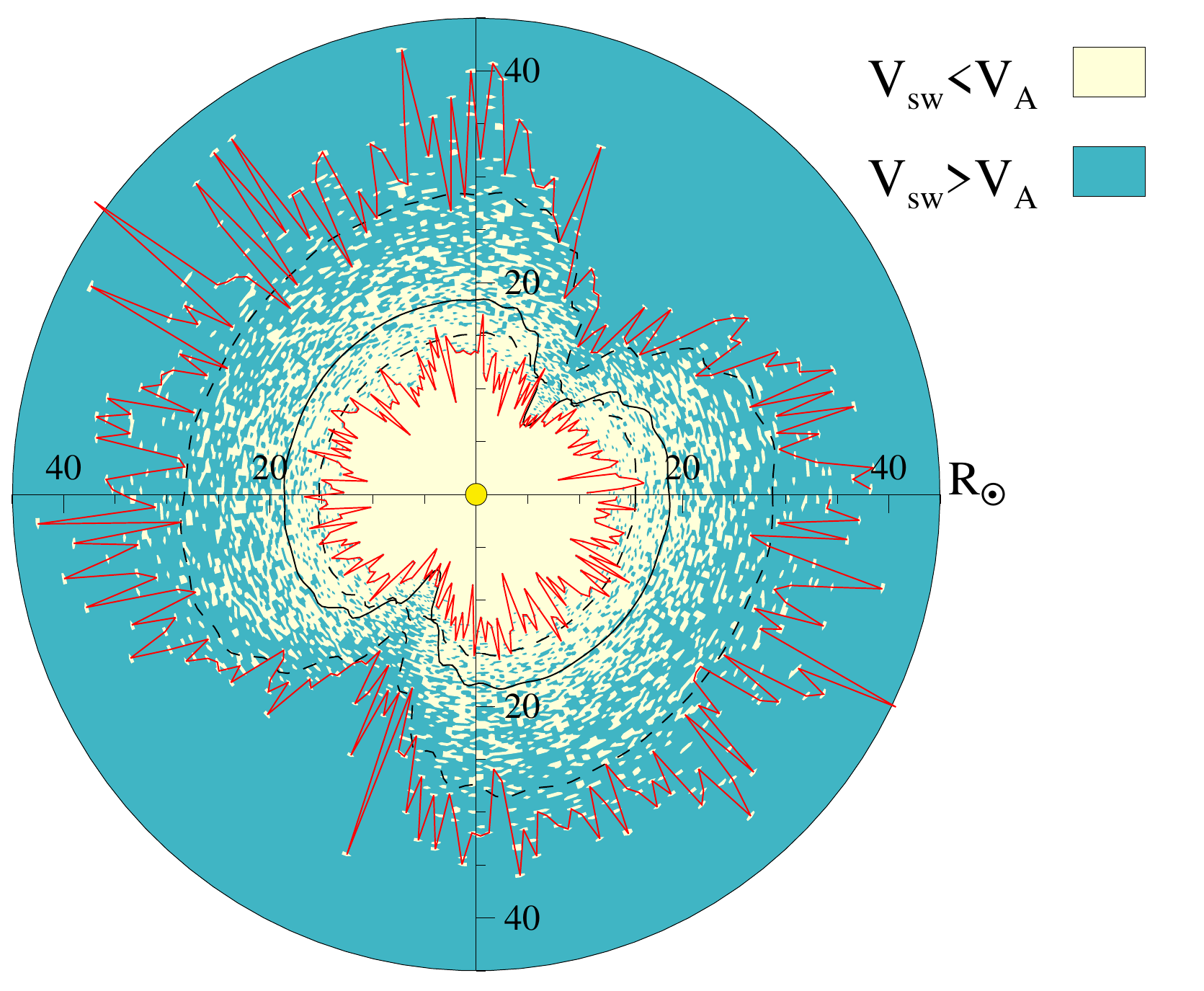}
    \includegraphics[width=1\columnwidth]{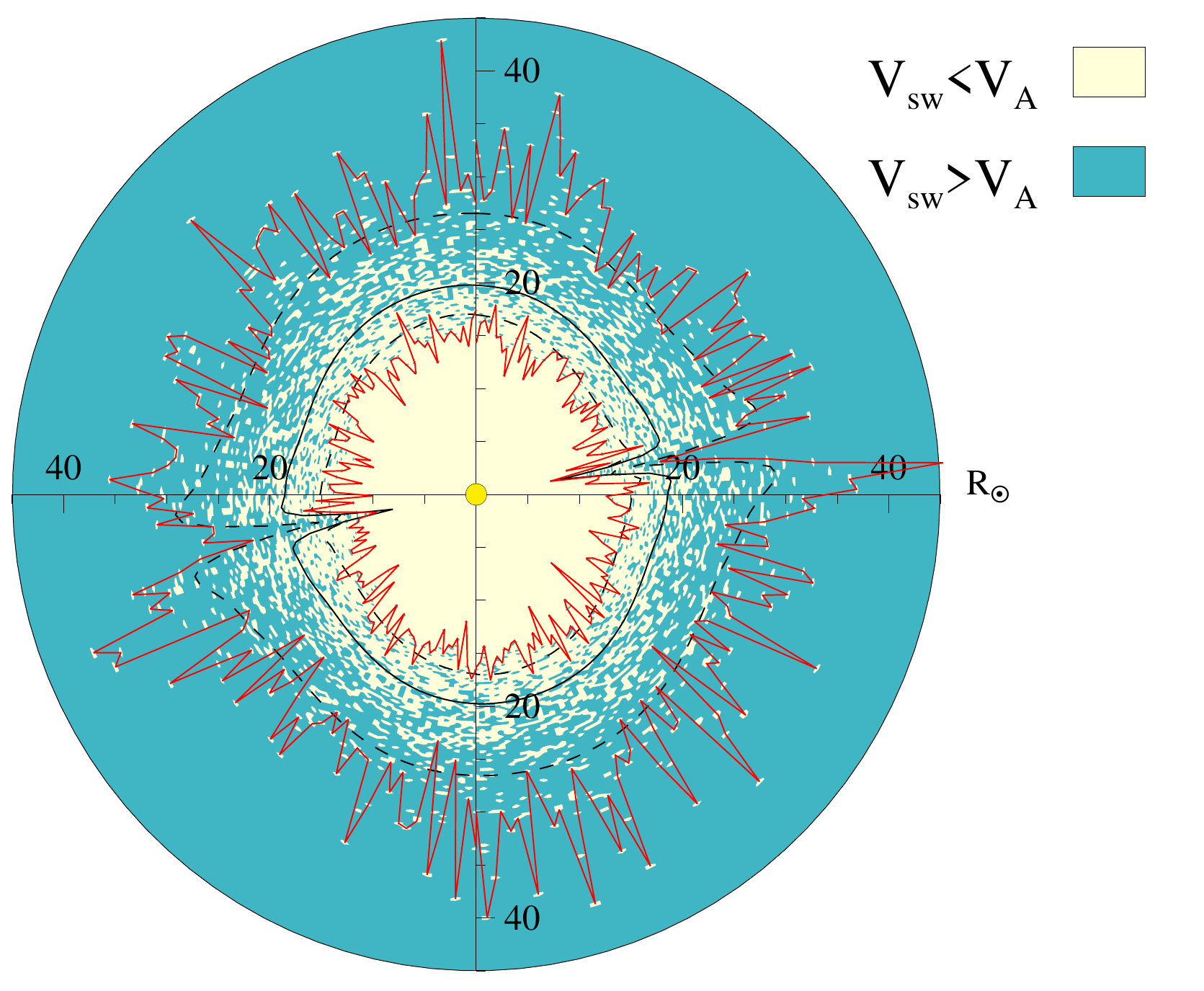}
    \caption{Sub-Alfv\'enic (beige) and super-Alfv\'enic (teal) regions are distinguished by colour, in a calculation that includes an explicit realization of magnetic turbulence (see text). Left and right panels show the solar equatorial plane and a meridional plane at 155\degree~longitude, respectively, for a \(10\degree\)-tilt dipole simulation (Run I). Helioradii ranging from 1 to \(45~\rs\)~are shown. Solid black curve shows the Alfv\'en surface computed from the mean fields. The two dashed black curves demarcate the ``average'' trans-Alfv\'enic region obtained by adding/subtracting the local rms magnetic fluctuation amplitude to the mean magnetic field, and correspond exactly to the boundaries of the red-shaded zone in Figure \ref{fig:stat_zone} (see Section \ref{sec:stat_zone}). The inner red curve marks, for each \(\theta\) and \(\phi\), the first super-Alfv\'enic (teal) point while moving outward along a radial spoke. Similarly, the outer red curve marks the last sub-Alfv\'enic (beige) point while moving outward along a radial spoke.}
    \label{fig:signum_dipole}
\end{figure*}

\begin{figure*}
    \centering
    \includegraphics[width=1\columnwidth]{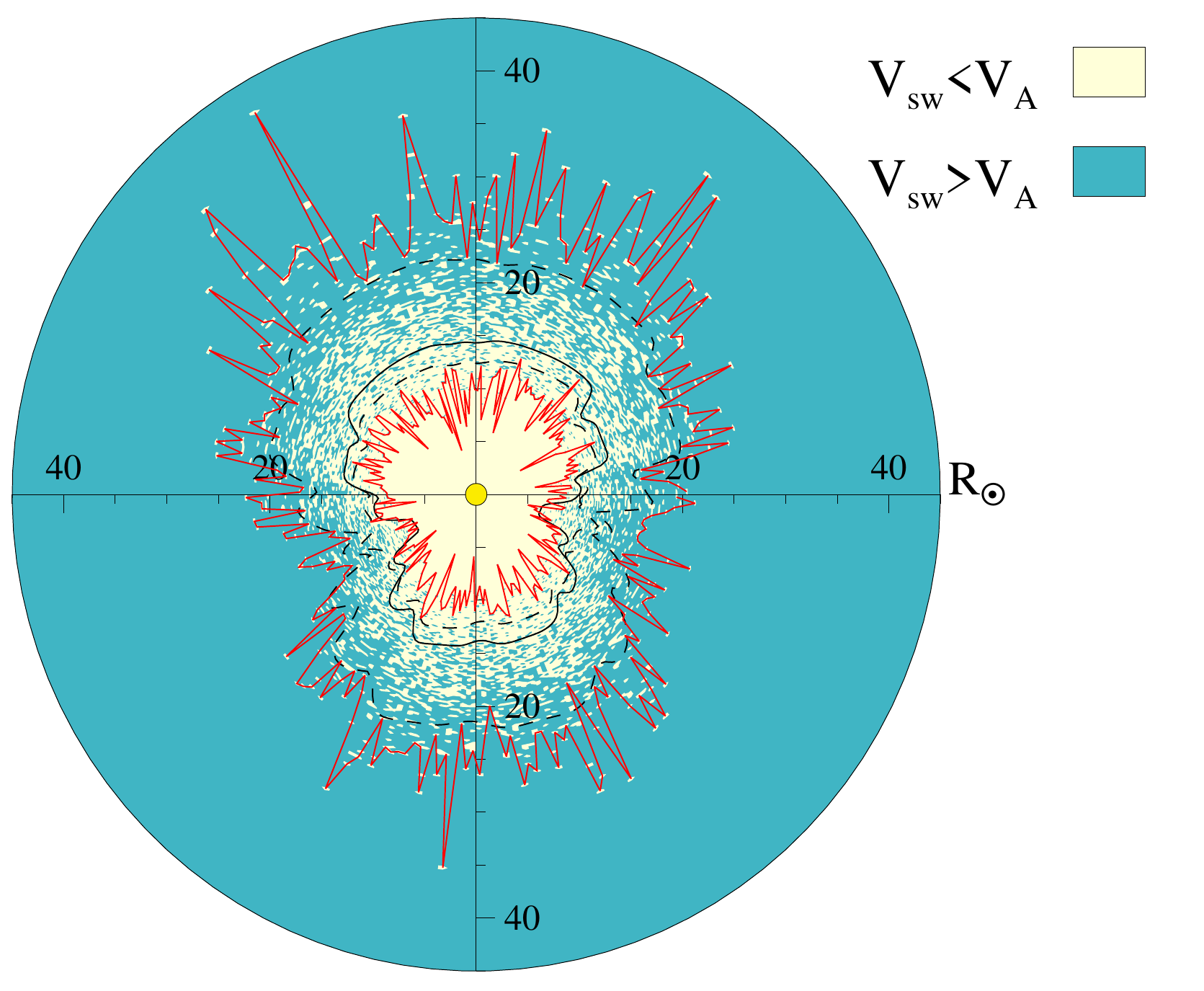}
    \includegraphics[width=1\columnwidth]{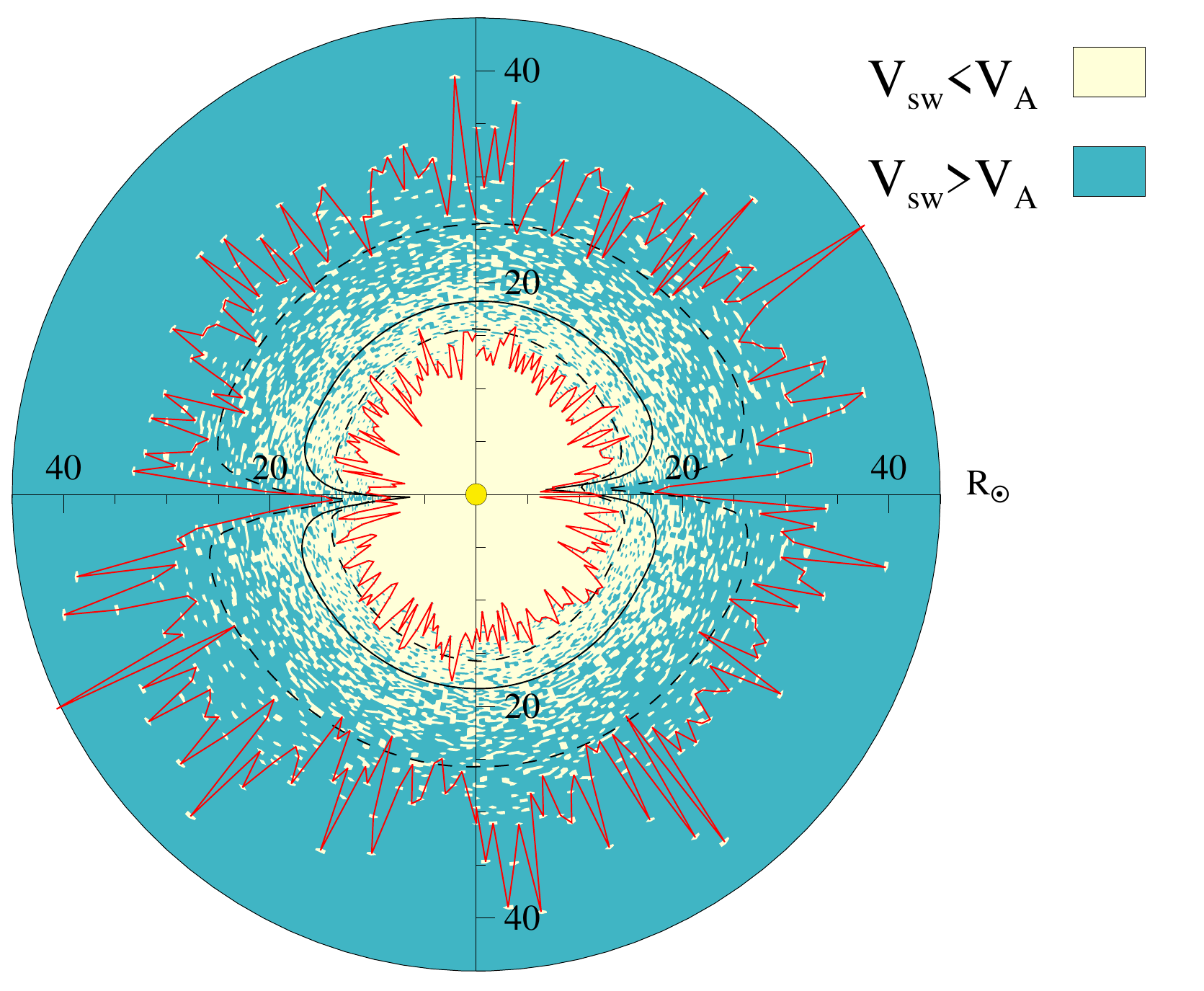}
    \caption{Left and right panels show the solar equatorial plane and a meridional plane at 155\degree~longitude, respectively, from Run II, based on a Nov 2018 magnetogram. The description follows Figure \ref{fig:signum_dipole}.}
    \label{fig:signum_mgram}
\end{figure*}

Figure \ref{fig:signum_dipole}
shows a further analysis of the tilted dipole run, 
including both an equatorial plane (left)
and a meridional plane (right).
Within  each, the super-Alfv\'enic (teal) and
sub-Alfv\'enic (beige) regions are distinguished by colour, where the realization of magnetic turbulence is included in
the determination as in Figure \ref{fig:MA_merid_10tilt}.
The solid black line (near 20 \(\rs\)) indicates the 
position of the simple critical surface that is obtained when the 
turbulence is not taken into account (i.e., only the 
large-scale mean fields are included in Equation \ref{eq:MA}), and corresponds exactly to the red curve in Figure \ref{fig:stat_zone}.
Note that if no turbulence were present the interior 
of the black contour would be entirely beige, and the exterior, teal. 
It is apparent that the effect of the 
turbulence is to cause a disjoint or fragmented 
transition between super- and sub-Alfv\'enic regions. Moving inward from large distances, one is likely to encounter
sporadic patches of sub-Alfv\'enic flow in regions that are dominantly super-Alfv\'enic.\footnote{The granularity of the patches is constrained by the resolution of the simulation grid. Recall that the grid resolution is of the order of a few correlation lengths, which is consistent with our approach in which random synthetic fluctuations at neighboring vertices are uncorrelated. Grid spacings finer than a correlation scale may require a different approach.} Below the black contour the dominance of the sub-Alfv\'enic patches increases and 
the patchiness eventually gives way to pure sub-Alfv\'enic wind.
The radius within which the flow is purely sub-Alfv\'enic
is indicated by the inner red contour. For this run that distance is irregular, but roughly near \(12~\rs\). 
Similarly, moving inward along any radial spoke,
the largest distance at which a sub-Alfv\'enic patch or blob is encountered, is indicated by the outer red contour. The two dashed black curves mark the ``average'' trans-Alfv\'enic region obtained by adding/subtracting the local rms magnetic fluctuation amplitude to the mean magnetic field, and correspond exactly to the boundaries of the red-shaded zone in Figure \ref{fig:stat_zone}.

Very similar properties are found when the above analyses are carried out using a simulation run initialized with 
boundary conditions derived from an appropriate 
magnetogram (see Section \ref{sec:runs}).
This approach affords a greater degree of realism
and a rough contextual connection
to conditions seen in PSP observations.
Figure \ref{fig:signum_mgram} shows results of such an analysis based on Run II, in the same format as the previous figure. 
In this case the Alfv\'en Mach number in the 
equatorial plane displays a somewhat more irregular pattern in comparison to the 
tilted dipole run in Figure \ref{fig:signum_dipole}.
This irregularity is manifest in both the overall  morphology including the turbulence realization, and is also seen in the black contour evaluated from the mean fields alone. The inner and outer contours (in red)
demarcating the domain of the Alfv\'en critical zone
are also rather irregular in his case, and are found closer to the sun than in the dipole case. This is partially due to the rapid radial decay of the higher-order multipole magnetic fields that are implied by a complex magnetogram boundary condition \citep{reville2015ApJ798,chhiber2019psp1}, and also due to the closer position of the HCS relative to the solar equatorial plane in the magnetogram. 
The latter point is made clear in 
the right panel of Figure \ref{fig:signum_mgram}, which illustrates the trans-Alfv\'enic region in a meridional plane using the same magnetogram run.

\begin{figure}
    \centering
    \includegraphics[width=1\columnwidth]{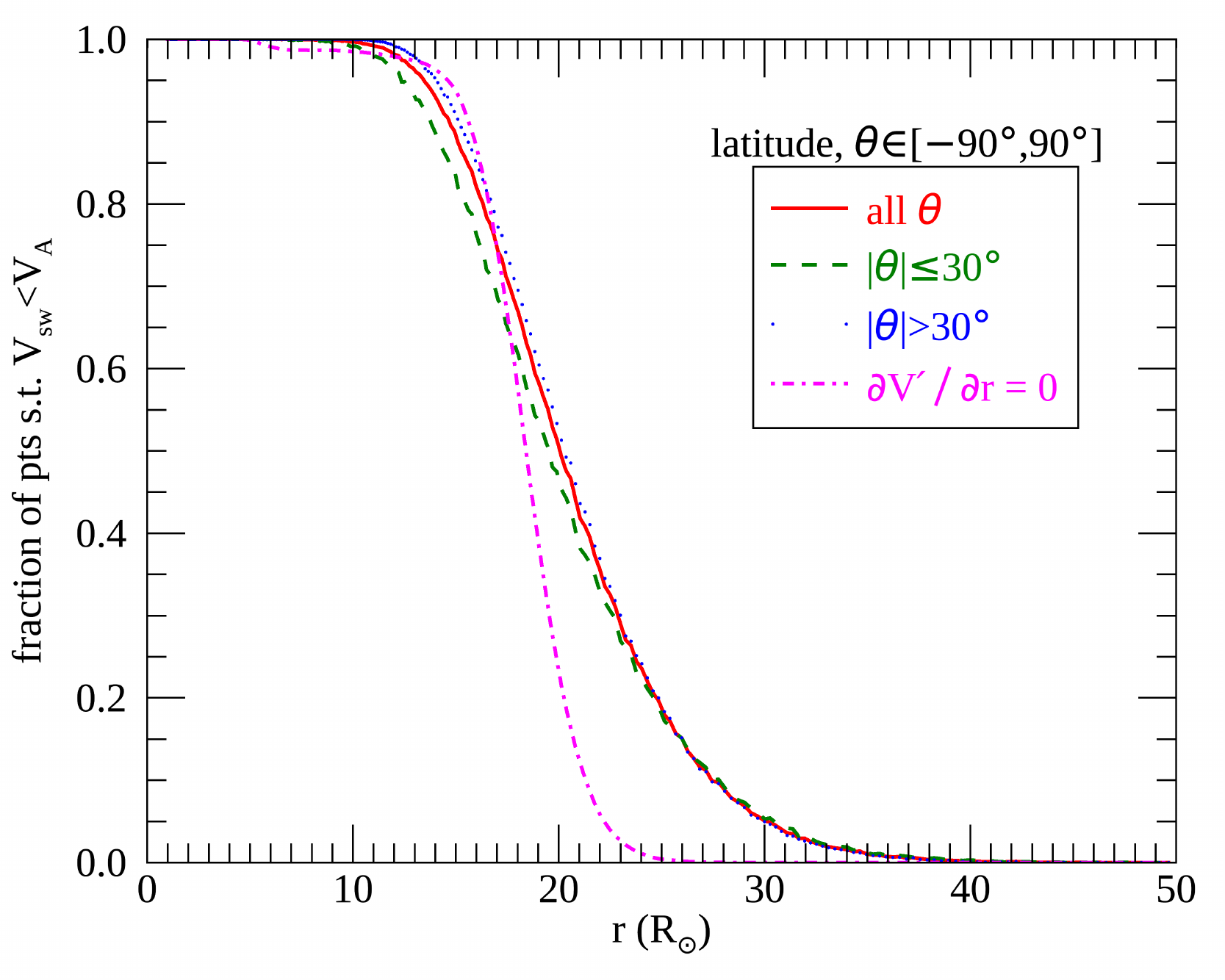}
    \caption{Filling fraction of sub-Alfv\'enic plasma, based on Run I. Fractions for each \(r\) are computed across all longitudes and indicated latitudes The solid red, dashed green, and dotted blue curves are based on the realization of magnetic fluctuations described in Figure \ref{fig:signum_dipole}. 
    The dash-dotted magenta curve is based on a realization of random velocity fluctuations (ignoring magnetic fluctuations) that do not vary in \(r\) and therefore produce a single ``wrinkled'', or corrugated Alfv\'en surface (see text). Fractions at each \(r\) for this curve are computed for all longitudes and latitudes.
    }
    \label{fig:frac}
\end{figure}

The panels of 
Figure \ref{fig:signum_mgram}
also confirm the main finding of this study -- that the transitions between sub-Alfv\'enic and super-Alfv\'enic 
flows occur initially in patches, which become more densely packed moving across the Alfv\'en transition zone. 
This gradual transition is 
readily quantified by computing the fraction of points
in the simulation that are sub-Alfv\'enic, 
as a function of radial distance.
For clarity, the results are averaged over solar 
longitude, and subject to a simple sorting by latitude. 
Figure \ref{fig:frac} shows the result of such 
calculations, for the tilted-dipole simulation discussed above (Run I). This is done for the full model including turbulence at high latitudes $|\theta| > 30^\circ$
and at low latitudes  $|\theta| < 30^\circ$, and for a simple model of a single ``wrinkled'', or corrugated Alfv\'en surface (see caption). \footnote{The latter model is based on a conception of fluctuations that are advected by the mean flow and do not vary in \(r\), therefore implying a single transition from sub-Alfv\'enic to super-Alfv\'enic flow along each radial spoke. Motivated by recent coronagraph observations \citep{deforest2018ApJ} that indicate velocity fluctuations of the order of 100 \(\kmps\) in the region 5-15 \(\rs\), we generate a realization of random velocity fluctuations for each \(\theta\) and \(\phi\) on the simulation grid, drawn from a Gaussian distribution with a standard deviation of 100 \(\kmps\). The total solar wind speed (including fluctuations) is computed in an analogous way to the procedure described in Section \ref{sec:frag} for the magnetic field. This speed is then compared with the Alfv\'en speed computed from the mean fields to produce the dash-dotted magenta curve in Figure \ref{fig:frac}.}
The relatively gradual 
transition in 
the patchy Alfv\'en zone 
case
may provide a basis for 
distinguishing 
it from the
single folded-surface case (dash-dotted curve)
which transitions more abruptly. We cannot rule out that varying the procedure used to model the latter case may modify the comparison in Figure \ref{fig:frac}.

\subsection{Comparison with PSP Observations}

The fields computed by the 
Reynolds-averaged simulations
have been recently compared with PSP observations in
some detail \citep{chhiber2021ApJ_psp}. 
In that case a number of parameters were found to agree well with PSP observations along its orbit,
and in some cases, 
the envelope of variations implied by the computed 
turbulence parameters demonstrated potential consistency 
between simulation and observation. For one case in that
paper, a realization of switchbacks 
was compared
with PSP magnetic field data to demonstrate both successes and shortcomings of this approach. 
Here we take this approach to comparison of PSP with simulation a step further and compare samples of the 
Alfv\'en Mach number time series from PSP data with 
a synthetic time series derived from 
a realization of the full 3D 
fields that determine the Mach number.  

\begin{figure}
    \centering
    \includegraphics[width=1\columnwidth]{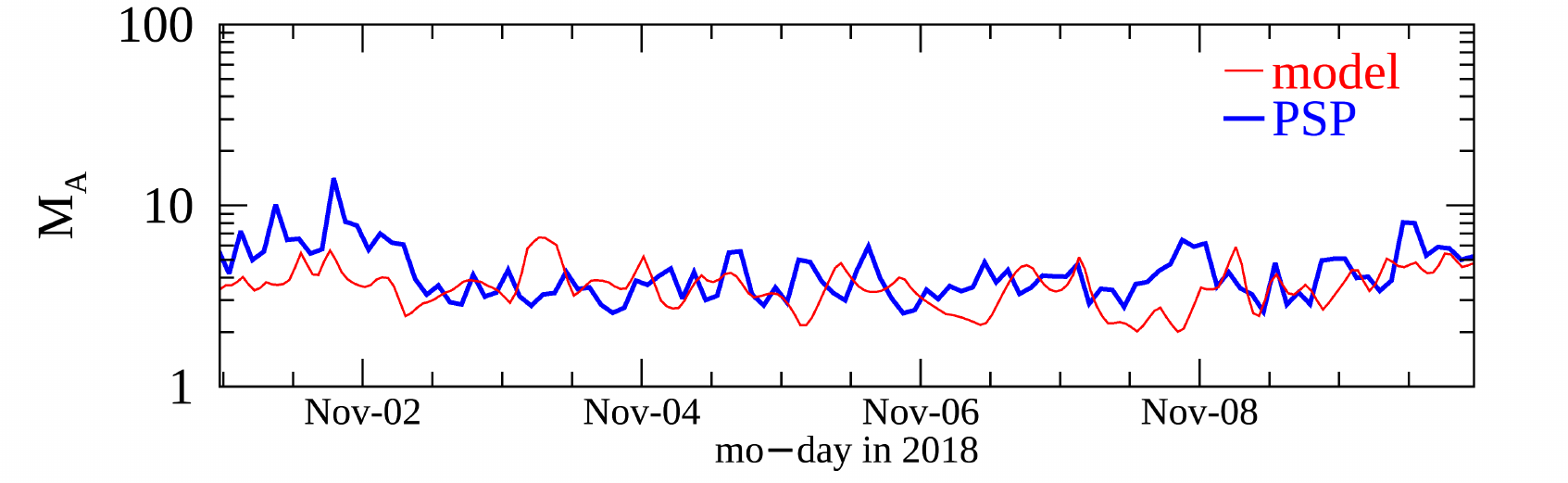}
    \includegraphics[width=1\columnwidth]{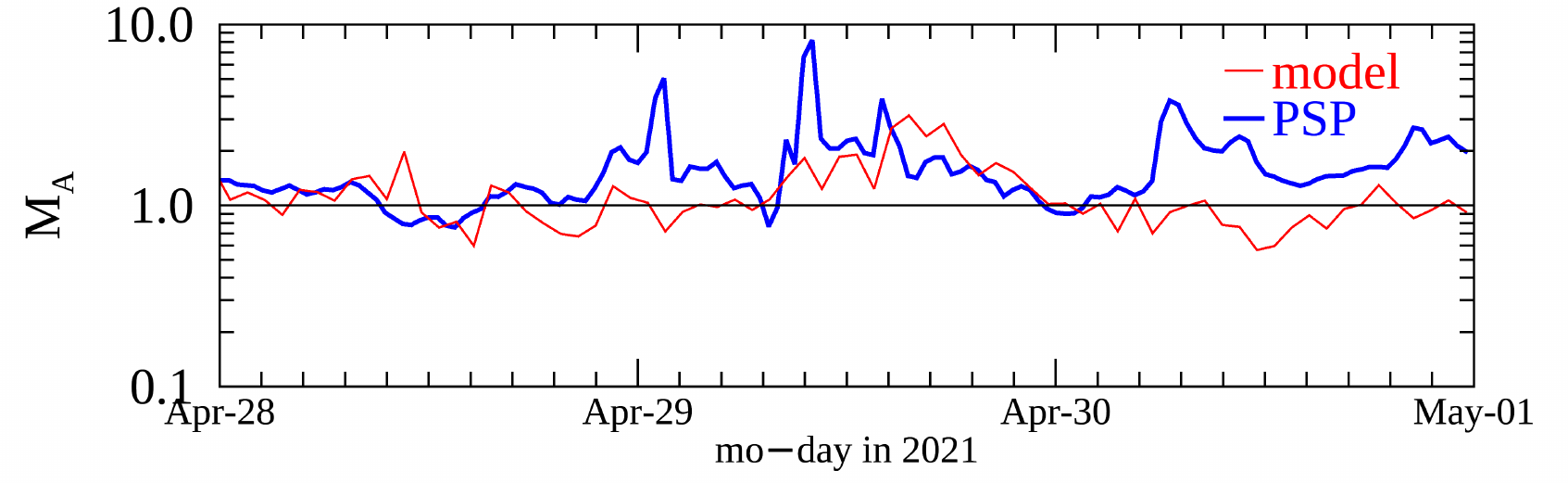}
    \caption{Alfv\'en Mach number along simulated PSP trajectory (thinner red curve), compared with PSP observations (thicker blue curve). Data are plotted at 1-hour cadence. Top (bottom) panel is based on Run II (Run III) and PSP observations from the first (eighth) solar encounter. The model curves include the effect of synthetic magnetic fluctuations, as described in Section \ref{sec:frag}.
    }
    \label{fig:PSP}
\end{figure}

Using the simulations driven by magnetograms corresponding to PSP's first and eighth encounters (Runs II and III, respectively), we construct the spatial distribution of mean fields and turbulence parameters, and then, a realization of 
the turbulence, which is superposed on the 
computed mean fields, as described in Section \ref{sec:frag}.
The Alfv\'en Mach number is computed throughout the simulation domains, and then (trilinearly) interpolated to the PSP trajectory at 1-hour cadence. The observations are smoothed to 1-hour cadence for comparison with the model. Two samples of results of this type are shown in Figure \ref{fig:PSP}. 

The top panel shows the Alfv\'en mach number \(\MA\) for an approximately eight day period at the beginning of November 2018, near the first perihelion at about \(36~\rs\) \citep[for details of PSP-data processing see][]{chhiber2021ApJ_psp}. The model result, including synthetic fluctuations, qualitatively matches the PSP data, in terms of general trend and in terms of the magnitude of the fluctuations. Of course, the comparison of actual waveform to synthetic waveform should not be taken seriously, since only the local variances are physically meaningful. 

The bottom panel of Figure \ref{fig:PSP}
shows another comparison of \(\MA\) of the same type,
in this case near the eighth perihelion at \(\sim 16~\rs\). For details on the processing of PSP measurements see \cite{BandyopadhyayEA21}. It is apparent that the Alfv\'en Mach number dips below unity several times in this period; these are the first sub-Alfv\'enic periods observed by PSP, as reported by 
\cite{kasper2021prl} and further analyzed by 
\cite{BandyopadhyayEA21}.
The agreement is again rather good considering that the phases of the fluctuations are random, and only the local variances are meaningful.  
For example, the more extended period of sub-Alfv\'enic flow in the second half of April 28 is approximately accounted for by model, but the sub-Alfv\'enic period seen only in the simulation between April 30 and May 1 is apparently an artifact.  

\section{Discussion}\label{sec:disc}

A number of current and 
recent observations \citep{liu2021ApJ,wexler2021ApJ} suggest that the Alfv\'en critical surface is not 
a simple smooth symmetric surface.
Remote sensing observations
\citep{lotova1985AA150,lotova1988SoPh117}
indicate a range 
of distances in which a critical transition might occur.
Recent detections of sub-Alfv\'enic regions near PSP perihelia also indicate a possible range of distances associated with the transition \citep{kasper2021prl}. Predictions of ranges of distances also have been 
reported based on extrapolation of in-situ measurements \citep{goelzer2014JGR,kasper2019ApJ, wexler2021ApJ}. In addition to a distorted or folded simple surface, more complex possibilities exist that amount to a state in which the transition occurs over a significant range of distances -- in an Alfv\'en transition zone  or region \citep{lotova1985AA150,matthaeus2004_alfven_region,deforest2018ApJ,chhiber2019psp1}. Based on the expected large amplitude of turbulent 
fluctuations near the critical region, we postulate that the transition occurs in fragmented patches rather than in a folded or distorted simply-connected surface. Here we provide 
empirical support for this possibility based
on a Reynolds-averaged model that includes, crucially, turbulence transport \citep{usmanov2018}. 

One feature we discussed in some detail is the extent
of 
the Alfv\'en critical region, 
bounding it with inner and outer 
surfaces defined by the first and last trans-Alfv\'enic blob along each radial ``spoke''. Between these lies the simple Alfv\'en surface that would be present if 
no turbulent fluctuations were present and the Alfv\'en mach number is calculated in terms of only the computed mean fields. 
Arguably, any of these three contours (inner, mean field, or outer) could be used to define an ``Alfv\'en surface'', depending
on the intent of the adopted definition. 
But it is equally clear
that for this realization there is no 
simply-connected critical surface
in the usual sense \citep{matthaeus2004_alfven_region}.
Instead, the above constructs provide a 
demarcation of what might 
properly be called 
a trans-Alfv\'enic zone.

A salient feature of this 
empirical demonstration of a fragmented 
Alfv\'en critical zone is the appearance 
of subvolumes or blobs of 
sub-Alfv\'enic (super-Alfv\'enic)  flow in regions that are predominantly super-Alfv\'enic (sub-Alfv\'enic). The blobs are of varying size and are visibly clumped in many locations, indicating correlations among the population. Their random character emerges as a convolution of the 
complex dynamics driven by the 
magnetogram boundary conditions, and the 
stochastic realization of the turbulence amplitudes
(implemented here in an {\it ad hoc} procedure constrained by the local value of the average turbulence amplitude provided by the model). 
The latter property is not entirely an artifact, since the random numbers employed are drawn for each computational vertex, which are separated by somewhat more than a turbulence correlation length. Therefore the vector fields in adjacent vertices are expected to be almost uncorrelated even if the turbulence amplitudes maintain a mutual correlation, 
established at the photospheric boundary and  maintained approximately along 
(many) nearby MHD characteristics. 

Another feature of the trans-Alfv\'enic
blobs is that their aggregate perimeter, or the length of their ``coastline''
is certain to be quite large. 
The question of whether the lengths of these perimeters scale, like the coastline
of Britain, in a way that signifies a fractal distribution \citep{mandelbrot1967Sci} is a technical question that we defer to future
study. In any case, 
the MHD wave characteristics
in this region are expected to 
be extremely complex
\citep{verdini2009ApJ},
and the behaviour of waves at these apparently complex interfaces
may also be an ingredient of coronal heating models worthy of additional examination. 
Finally, we might ask whether the irregular 
distribution of trans-Alfv\'enic patches or blobs represent a type of spatial intermittency.
Referring again to the relatively large size of these patches (in the present implementation),
it would seem doubtful that their distribution 
contributes directly to either inertial range or dissipation range intermittency
\citep{sreenivasan1997AnRFM,matthaeus2015ptrs},
which are generally described in 
terms of increments at 
lags much smaller than the correlation scale
\citep[e.g.,][]{sorriso-valvo1999GRL,Chhiber2021ApJL}.
However, the type of large-scale intermittency described by
Oboukhov in his seminal 
1962 paper may be relevant here: in particular,
the author writes
``These slow fluctuations of energy dissipation are due to change of the large-scale
processes in the observation region, or 
`weather' in a general sense.'' \citep{oboukhov1962JFM}.
This directly motivates Oboukhov's introduction of log-normal statistics into the formulation 
of turbulent intermittency. Lognormal statistics are pervasive in solar wind parameters 
including the distributions of magnetic field strength \citep{burlaga1998JGR} and correlation scale 
\citep{Ruiz2014SoPh}, and is a factor underlying the possible origin of so-called \(1/f\) noise in the interplanetary magnetic field 
\citep{matthaeus1986prl}. The distribution
of fragmented trans-Alfv\'enic regions is still
another possible example of large-scale intermittency due to gusty solar ``weather''. Indeed there 
may be many interesting connections and directions for further work indicated by the current conceptual and 
semi-empirical study.

One can imagine reasonable modifications to the procedure used here to generate synthetic fluctuations; these include different values for the Alfv\'en ratio, variance polarization anisotropy, and the inclusion of velocity and density fluctuations in addition to magnetic. Nevertheless, we do not expect such changes to affect the qualitative conceptual picture we have proposed. We emphasize the novelty of our approach for generating an explicit realization of turbulent fluctuations within the context of a Reynolds-averaged global solar wind model; with further refinements this technique can find a variety of applications, such as studies of energetic particle transport \citep[e.g.,][]{moradi2019ApJ,chhiber2021AA}.

The observational question that remains is 
whether the simple surface model or the 
extended, fragmented zone model is more realistic. Our empirical model provides 
predictions for future PSP observations, which 
might possibly be used to distinguish between ``surface vs zone'' pictures -- our model implies greater density/frequency-of-occurrence of sub-Alfv\'enic patches, and longer durations as well, as PSP descends to lower perihelia. 
The surface picture implies longer durations of sub-Alfv\'enic intervals as PSP descends, instead of increasing density and frequency of such intervals; this picture also most likely is associated with sharper transitions (see Figure \ref{fig:frac}). 

Later PSP orbits will provide opportunities for possible support for the present perspective.  
We would argue that successive orbits at lower heliocentric distances will continue to observe patches of sub-Alfv\'enic flow but at increasing rates and higher filling fractions at lower perihelia, until finally arriving at pure or nearly pure sub-Alfv\'enic flow. There is no guarantee that PSP orbits will probe deeply enough to arrive in the pure sub-Alfv\'enic coronal plasma. 
Even still, the radial trends may permit further analysis to distinguish the fragmented patchy transition
that we explore here, from a simpler 
``wrinkled'', or corrugated surface model such as that we described above. This distinction may prove to be important in distinguishing different coronal heating models, since a fragmented transition zone may provide a kind of ``pressure
cooker'' for enhanced heating by interaction of 
reflected ``inward'' type waves with the dominant outward propagating Alfv\'enic fluctuations \citep{matthaeus1999ApJL523}. Other types of simulations, such as the so-called expanding box model \citep{grappin1993prl,squire2020ApJ}, may also provide additional insights into the ``surface vs zone'' question.
In any case, observations by any single spacecraft 
may not be able to recover enough 3D 
information to conclusively determine the topography of the Alfv\'enic transition zone. 
It is, however, hoped that sophisticated polarized imaging instruments on the PUNCH mission
\citep{deforest2019AGU_PUNCH}, designed to provide high resolution
3D data, will be able to make this important determination. 

\section*{Acknowledgements}
This research partially supported by NASA under the Heliospheric Supporting Research program grants
80NSSC18K1210 and 80NSSC18K1648 and by the Parker Solar Probe Guest Investigator program 80NSSC21K1765, the 
PSP/IS\(\odot\)IS\ Theory and Modeling project (Princeton 
subcontract SUB0000165), and the PUNCH project under subcontract 
NASA/SWRI N99054DS.
We thank Dr. Francesco Pucci for suggesting the designation
``antarctic plot'' for Figures 
\ref{fig:signum_dipole}
and 
\ref{fig:signum_mgram}.
This work utilizes data produced collaboratively between AFRL/ADAPT and NSO/NISP.
Computing resources supporting this work were provided by the NASA High-End
Computing (HEC) Program (awards SMD-17-1580 and SMD-17-1617) through the
NASA Advanced Supercomputing Division at Ames Research Center. We acknowledge the \psp\ mission for use of the data, which are publicly available at the \href{https://spdf.gsfc.nasa.gov/}{NASA Space Physics Data Facility}.

\section*{Data Availability}

Simulation data will be made available upon reasonable request to the corresponding author. PSP data are publicly available at the \href{https://spdf.gsfc.nasa.gov/}{NASA Space Physics Data Facility}.
 



\bibliographystyle{mnras}
\bibliography{chhibref}







\bsp	
\label{lastpage}
\end{document}